\documentstyle[multicol,aps,epsf,prb]{revtex} 

\newcommand{\be}{\begin{equation}}
\newcommand{\ee}{\end{equation}}
\newcommand{\bea}{\begin{eqnarray}}             
\newcommand{\eea}{\end{eqnarray}}
\newcommand{\nn}{\nonumber}
\newcommand{\bm}[1]{\mbox{\boldmath${#1}$}}
\newcommand{\grad}{\bm{\nabla}}
  
\begin{document}

\title{Physics on the edge: contour dynamics, waves and solitons
	in the quantum Hall effect}
\author{C. Wexler and Alan T. Dorsey}
\address{Department of Physics, Box 118440, University of Florida,
  Gainesville, Florida  32611-8440} 
\date{May 1999, modified July 1999}
\draft
\maketitle

\begin{abstract}
We present a theoretical study of the excitations on the edge of a
two-dimensional electron system in a perpendicular magnetic field in
terms of a contour dynamics formalism. In particular, we focus on edge
excitations in the quantum Hall effect.  Beyond the usual linear
approximation, a non-linear analysis of the shape deformations of an
incompressible droplet yields soliton solutions which correspond to
shapes that propagate without distortion.  A perturbative analysis is
used and the results are compared to analogous systems, like vortex
patches in ideal hydrodynamics.  Under a local induction approximation
we find that the contour dynamics is described by a non-linear partial
differential equation for the curvature: the modified Korteweg-de
Vries equation.  
\end{abstract}
%
%

\begin{multicols}{2}

\section{Introduction} \vspace{-.3cm}
\label{sec:intro}

The theoretical description of many-body systems is often best
realized in terms of collective modes, i.e. the familiar sound waves
in solids or plasmons in charged systems.  Collective modes
become especially important when their energies are lower than
competing single-particle excitations.  Sometimes, however, both
single-particle and collective modes in the bulk of a 
system are gapped or scarce and these systems are often referred to as
``incompressible.''  This incompressibility can be real or a
convenient limit due to large differences in 
relevant length- or time-scales, as in the case of the macroscopic
motion of a liquid.  Under these conditions, one can usually focus our
attention on the motion of the {\em boundaries} of the system, which will
generally have softer modes, with frequencies much lower than those in
the bulk (e.g. surface waves in a liquid droplet travel at speeds
considerably slower than sound waves).

Concentrating on the motion of the boundary of the
system has a considerable advantage: the reduction in the
dimensionality of the problem often permits simpler analytical
treatment, or a tremendous reduction in the effort needed to
numerically solve or simulate the problem.  Associated with this
incompressibility, however, one usually finds microscopic 
conservation laws that translate into global constraints on the whole
system, even when the microscopic dynamics is completely local
(e.g. the volume of a liquid droplet is conserved).
These global constraints enter the edge dynamics through Lagrange
multipliers associated with the conserved quantities.\cite{mechanics}
These conservation laws are often evident when the motion of the
system is observed, and it is interesting to see how they are embedded
in the dynamics of the boundary; that is, how these essential aspects of
the problem are related to the laws of motion of the edges.

These {\em shape deformations}, and their dynamics, have played an
important role in the understanding of numerous problems in
diverse fields of physics.  The incompressibility is reflected in the
existence of a field which is piecewise constant, so that there is a
sharp boundary between two or more distinct regions of space with
different physical properties.  This field can be of classical origin
like the density of a liquid or the charge density of a plasma, or it
can originate in the quantum mechanical properties of the system, 
like the magnetization of a type-II superconductor. 
There are various examples where these questions are relevant, such as
waves on the surface of a
liquid,\cite{LLFM,LiquidDropEXPT,LiquidDropMKDV} the motion of 
non-neutral plasmas,\cite{non-neutral} low lying
``rotation-vibration'' modes of deformed nuclei,\cite{Nuclear}
the evolution of atmospheric plasma clouds,\cite{PlasmaClouds}
pattern formation in ferromagnetic fluids,\cite{PatternFormation}
vortex patches in ideal
fluids,\cite{lamb,ContourAlgorithm,VStates,VortexPatchDynamics} and 
two-dimensional electron systems (2DES) in strong magnetic
fields.\cite{EdgesQHE,wexler99}

The edges of a two-dimensional electron system (2DES), and in
particular the edges of a quantum Hall (QH) liquid, present a unique
opportunity to study the dynamics of shape deformations in a clean and
controlled environment.  The 2DES in the QH state is incompressible, so
that the electron density is approximately piecewise constant, 
suggesting that a contour dynamics approach to studying the droplet
excitations is viable.  The lack of low-lying excitations eliminates
dissipative effects, further simplifying the treatment of the problem. 
In addition, the charged nature of the system facilitates the
excitation and detection of deformations of the droplet. 

In this paper we will formulate the study of the excitations of a
droplet in a 2DES as a problem in contour dynamics.  In the usual
treatment of the edge excitations,\cite{EdgesQHE} a linearization of
the equation of motion is done at early stages, thus limiting the
applicability to small deformations of the edge of the system from an
unperturbed state. In this paper we consider non-linear terms which
are present in the full contour dynamics treatment. We first present
perturbative results for non-linear deformations of the 2DES
shape. For the sake of comparison, and as means of verification, we
also apply this method to the vortex patch case, which has well known
exact \cite{lamb} and numerical \cite{VStates} solutions. We
then show that the local induction approximation to the full contour
dynamics generates the modified Korteweg-de Vries (mKdV) \cite{mKdV}
equation for the curvature dynamics; the mKdV equation also arises in
studies of vortex patches \cite{VortexPatchDynamics} and suspended
liquid droplets.\cite{LiquidDropMKDV}  The mKdV dynamics conserve an
infinite number of quantities, including the area, center of mass, 
and angular momentum of the droplet,\cite{KdVHierarchy}
so that our local approximation to the nonlocal dynamics preserves the
important conservation laws. The mKdV equation also possesses soliton
solutions, including traveling wave solutions.

In Sec.\ \ref{sec:hydro2DES} we present a brief review of the
hydrodynamics of a two-dimensional electron system in a strong
perpendicular magnetic field and analyze the bulk and edge excitations
in the linear approximation.  Sections \ref{sec:dyn} and \ref{sec:nep}
and analyze in more detail the dynamics and kinematics of these edge
modes, and ask what conditions must be satisfied so that a large edge
deformation is able to travel without any dispersion, that is,
preserving its shape.  The question is posed in terms of a non-linear
eigenvalue problem and is solved perturbatively to fifth order in the
deformation in Sec.\ \ref{sec:solving-nep} (for completeness, we also
solve the analogous problem of vortex patch deformations in Appendix
\ref{sec:vortex-patch}).  Some solutions and limiting cases are then
presented in Sec.\ \ref{sec:solving-nep-2DES}. An alternative approach
to find non-dispersive or invariant deformations of the edge, namely
the {\em local induction approximation}, is developed in Sec.\
\ref{sec:lia}, where results are also compared with the perturbative
approach of Secs.\ \ref{sec:solving-nep} and
\ref{sec:solving-nep-2DES}. 


\section{Hydrodynamics of a two-dimensional electron system}  \vspace{-.3cm}
\label{sec:hydro2DES}

Consider a 2DES in a perpendicular magnetic field. Treated as
a classical fluid, the system is characterized by the electron density
$n({\bf r})$ and velocity ${\bf v}({\bf r})$. If we neglect
dissipative effects (which is essentially correct in the quantum Hall
regime), the dynamics is determined by the Euler and continuity
equations:
\bea
\label{eq:euler1}
\dot{\bf v} &\equiv&
   \frac{\partial {\bf v}}{\partial t} + ({\bf v \cdot \grad}) {\bf v}
 \nn \\
&  =& -\omega_c \, {\bf e}_z \!\times\! {\bf v} 
- \frac{e^2}{m_e \epsilon} \grad \!\!
\int d^2r' \, \frac{n({\bf r'})}{|{\bf r}-{\bf r'}|} 
+ \frac{e}{m_e} {\bf E}_{\rm ext}  \;, \\
\label{euler2}
&&{\partial_t n} + \mbox{\boldmath$\nabla$} \!\cdot\! 
        ( n {\bf v}) = 0 \,, 
\eea
where $\omega_c \!=\! e B /m_e$ is the cyclotron frequency and
$\epsilon$ is the dielectric constant of the medium. The first term on
the right-hand side of Eq.\ (\ref{eq:euler1}) represents the Lorentz
force,  the second is the Coulomb interaction, and ${\bf E}_{\rm ext}$
is the electric field due to the background positive charge, gates,
etc.

Consider first the possible bulk collective excitations in a uniform
system. These are oscillations of the density around the equilibrium
solution ${\bf v}\!=\!0$, $n\!=\!\bar{n}$. If we consider small
amplitudes and Fourier analyze Eqs.\ (\ref{eq:euler1}) and
(\ref{euler2}), we find that 
\bea
&& \!\!\! \left\{ 
        \begin{array}{l}  \displaystyle
        {\bf v}({\bf r},t) = {\bf v}_0 \, 
        e^{i {\bf k \cdot r} - i \omega t} \,,\\
\displaystyle   \delta n ({\bf r},t) = \frac{\bar{n}}{\omega} \,
        {\bf k \cdot} {\bf v}_0 \, e^{i {\bf k \cdot r} - i \omega t}\,,
        \end{array} \right. \\
&& \omega^2 = \omega_c^2 + \frac{2 \pi \bar{n} e^2}{m_e \epsilon} k \,,
\eea
that is, bulk magnetoplasmons are gapped, and in the QH regime can be
neglected since $\hbar \omega_c \!\sim\! 17 {\rm meV} \!\sim\! 200
{\rm K}$ at B = 10 T (in GaAs), whereas T $\sim$ 1K.  We will disregard
them from now on and concentrate on the excitations at the edge which,
as we shall see, have considerably lower energies.

The theory of small deformations of the edge has been extensively
studied.\cite{EMP,linear,winf,glazman}  The main conclusion is that 
for strong magnetic fields, when Landau level quantization 
becomes important, the only low-lying modes are edge modes which
propagate in only one direction along the edge of the 2DES. We further
classify these modes into the ``conventional'' edge magnetoplasmon
mode, with a singular dispersion relation:\cite{EMP}
\be
\label{eq:omega0}
\omega_0 (k) = -2 \ln \left( \frac{e^{-\gamma}}{2 |k a|} \right) 
   \frac{\bar{n} e^2}{\epsilon m_e \omega_c} k \,,
\ee
where $k$ is the mode wave-number, $\gamma \! \approx \! 0.5772$ is
the Euler constant, and $a$ is a short-distance cut-off
(the largest of the transverse width of the 2DES, the magnetic length,
or the width of the compressible edge-channel). In addition, for wide 
compressible edges, ``acoustic'' modes can be found: 
\be
\omega_j (k) = - \frac{2 \, \bar{n} e^2}{\epsilon m_e \omega_c \, j} k \,, 
\;\;\;\;\; j = 1,2,... 
\ee
These results are approximately correct as long as inertial and
confining terms are negligible. The first requires that 
$\bar{n} e^2/\epsilon m_e a \!\ll\! \omega_c^2$, which is usually true in
quantum Hall conditions.\cite{glazman}  In addition, the effects of
the confining potential have been neglected.  While the confining
potential is essential for long-term stability and is usually not
negligible when compared to interactions, its effect is mainly
reduced, in the simplest cases, to an additional $v_{\rm ext} \, k$
term in the frequencies, where the external velocity is given by
\be
\label{eq:vext}
v_{\rm ext} = - \frac{e}{m_e\omega_c} \, {\bf e}_z 
        \times {\bf E}_{\rm ext} \,.
\ee
Recent time-of-flight measurements \cite{AcousticModes} 
in 2DES confirm this simple picture. The theoretical
formulation above, however, is restricted to small deformations of the
boundary. In what follows we will consider a formulation that goes beyond
this limitation.

\section{Dynamics of the edge modes}  \vspace{-.3cm}
\label{sec:dyn}

\begin{figure}
  \begin{center}
    \leavevmode
    \epsffile{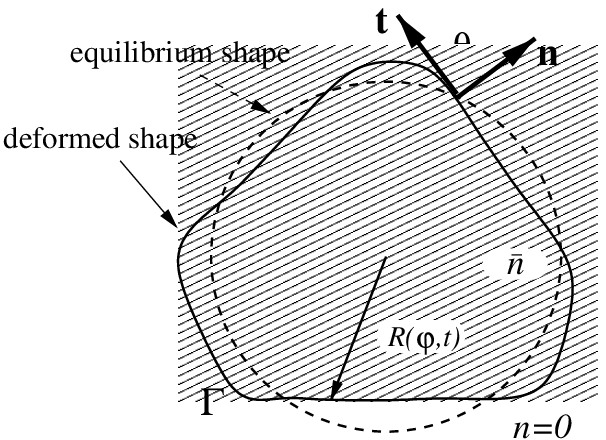}
  \end{center}
  \caption{\label{fig:droplet}
    A charged incompressible liquid in a magnetic field. We assume a
    piecewise constant electron density ($n\!=\! \bar{n}$ inside, while
    $n \!=\! 0$ outside the droplet). The parameterization
    $R(\varphi,t)$, the tangent ${\bf t}$ and normal ${\bf n}$ unit
    vectors to the boundary $\Gamma$ are indicated; $s$ is the
    arc-length and $\theta$ the tangent angle. }  
\end{figure}

Consider the case when the edge between bulk and vacuum
is sharp. The electronic density has some constant value $\bar{n}$
inside the edge and vanishes outside (see Fig. \ref{fig:droplet}).
Since the edge is essentially one-dimensional only the edge
magnetoplasmon is important (and even in more general cases this mode
is the most readily observable\cite{EMP,AcousticModes}).  Following
the derivation of the edge magnetoplasmons\cite{EMP,glazman} we
neglect inertial terms in Eq.\ (\ref{eq:euler1}), thus obtaining an
equation for the electron {\em velocity}:
\begin{equation}
\label{eq:vbounbd1}
{\bf v({\bf r})} = - \frac{e^2}{\epsilon m_e \omega_c}
\mbox{\boldmath$\nabla$} \times  {\bf e}_{z}
        \int_{\cal A} d^2r' \; \frac{n({\bf r'})}{|{\bf r}-{\bf r'}|} + 
        {\bf v}_{\rm ext} \,,
\end{equation}
where ${\cal A}$ is the area of the droplet (see Fig. \ref{fig:droplet}). 

Let us now concentrate on the ``internal'' velocity given by the first
term in Eq.\ (\ref{eq:vbounbd1}). We should note that the neglected
external velocity ${\bf v}_{\rm ext}$ derived from the confining field
[Eq.\ (\ref{eq:vext})] is important for long term stability of the
system, and will modify the propagation velocity.  In general, it
could also change the shape of the modes, yet one can devise
situations in which this effect is not important, i.e. a linear
confining potential in a rectilinear infinite edge, or a parabolic
potential for a circular droplet.

For an incompressible 2DES with a piecewise constant density, the
density can be taken outside the integral; then using Stokes' theorem,
the area integral can be transformed into a line integral over the
boundary $\Gamma \!=\! \partial {\cal A}$ of the electron liquid: 
\begin{equation}
\label{eq:vlineint}
{\bf v({\bf r})} = \frac{\bar{n} e^2}{\epsilon m_e \omega_c} 
        \oint_\Gamma ds' \,\frac{ {\bf t}(s')}{|{\bf r}-{\bf r}(s')|} \,.
\end{equation}
Here ${\bf t}(s')$ is the unit tangent vector at arc-length $s'$:
\be
\begin{array}{l}
        {\bf t}(s') = \partial_{s'} {\bf r}(s') \equiv {\bf r}_{s'} \,,\\
        {\bf n}(s') \equiv - {\bf e}_z \times {\bf t}(s') \,,
        \end{array} 
\ee
and we defined the unit normal vector ${\bf n}(s')$ for later use.  The
short distance singularity in the integrand is cut off at a length scale
$r_0$. Equation (\ref{eq:vlineint}) forms the basis of our contour
dynamics treatment---it expresses the velocity of the edge in terms of
a  nonlocal self-interaction of the edge. 

We see that: (i) the dynamics is chiral, being determined by the tangent
vector; (ii) the fluid contained within $\Gamma$ is incompressible, so
that the area  is conserved.  It is interesting to note similar traits
were found in the past for the dynamics of vortex
patches;\cite{lamb,ContourAlgorithm,VStates,VortexPatchDynamics}  
indeed, it is this analogy which inspired the present work.  A detailed
description of the vortex patch case, and stressing the similarities
and differences with the shape deformations of the 2DES is presented
for completeness in Appendix \ref{sec:vortex-patch}.

\section{Kinematics of uniformly propagating deformations}  \vspace{-.3cm}
\label{sec:nep}

Having determined the velocity of the electron liquid, we now focus on
the motion of the 2DES boundary $\Gamma$. The velocity of a point on the
boundary can be written in terms of the normal and tangential components, 
\begin{equation}
{\bf v} = U(s)\, {\bf n} + W(s)\, {\bf t}. 
\end{equation}
The tangential velocity $W(s)$ is largely irrelevant, as it solely
accounts for a reparametrization of the curve; the boundary motion is
determined by the normal component of the velocity.

\begin{figure}
\begin{center}
\leavevmode
\epsfbox{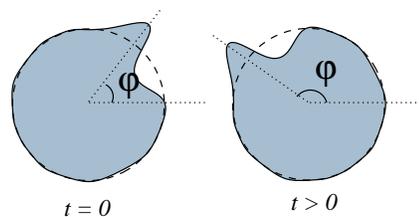}
\end{center}
\caption{\label{fig:invariant}
  A uniformly propagating edge deformation.  The boundary satisfies 
  $R (\varphi,t) = R (\varphi - \Omega t)$. }
\end{figure}

We now ask whether there are modes which propagate along the boundary
with no change in shape. Previous work \cite{EMP} has focused on small
perturbations of a straight, infinite edge.  Here we consider
deformations of a circular droplet of incompressible electrons (the
non-linear deformation of a straight infinite edge is discussed in
Appendix \ref{sec:infinite}). A uniformly propagating mode is, in this
case, characterized by a boundary that moves like a rotating rigid
body (see Fig. \ref{fig:invariant}), namely, the radius of the
boundary satisfies 
\be
R (\varphi,t) = R (\varphi - \Omega t) \,,
\ee
where $\varphi$ is the azimuthal angle, and $\Omega$ is the angular
frequency of the boundary rotation.  This translates into a condition
for the {\em normal} velocity: 
\begin{equation}
\label{eq:rigidrotation}
U \equiv \left. {\bf n}({\bf r})\cdot{\bf v}({\bf r}) 
        \right|_{{\bf r} \in \Gamma} = 
\Omega \; {\bf n}({\bf r})\cdot \left( {\bf e}_z \times {\bf r}\right)
\,.
\end{equation}

We seek surface shapes $R(\varphi)$ (see Figs.\ \ref{fig:droplet} and
\ref{fig:invariant}) that rotate uniformly, satisfying
Eq.\ (\ref{eq:rigidrotation}).  Consider the following parameterization
of the surface: 
\begin{equation}
\label{eq:paramet}
R(\varphi) = R_0 \left( 1 + \sum_{l=-\infty}^{\infty} b_l \; 
        e^{i l \varphi} \right) .
\end{equation}
In this parametrization, we can write the unit tangent vector
explicitly as
\be
{\bf t}(\varphi) = \frac{\bm{\tau}(\varphi)}{|\bm{\tau}(\varphi)|} \,,
\ee
where
\be
\label{eq:tau_phi}
\bm{\tau}(\varphi) 
        \equiv \frac{\partial {\bf r}(\varphi)}{\partial \varphi}
        = {\bf e}_r R'(\varphi) + {\bf e_\varphi} R(\varphi) \,.
\ee
Likewise, the unit normal vector is given by 
\be
\label{eq:n_phi}
{\bf n}(\varphi) = - {\bf e}_z \times 
\frac{\bm{\tau}(\varphi)}{|\bm{\tau}(\varphi)|} \,.
\ee
Given the identity ${\bf r}_s \, ds \!=\! {\bf r}_\varphi \, d\varphi$,
the normal velocity of the boundary, derived from Eq.\
(\ref{eq:vlineint}), can be written as
\begin{equation}
\label{eq:normalv}
U (\varphi) = \frac{\bar{n} e^2}{\epsilon m_e \omega_c} 
        \int_0^{2 \pi} \! d\varphi' \; 
        \frac{{\bf n}(\varphi) \cdot \mbox{\boldmath$\tau$}(\varphi') }
                {|{\bf R}(\varphi) - {\bf R}(\varphi')|} \,,
\end{equation}
while inserting Eqs.\ (\ref{eq:tau_phi}) and (\ref{eq:n_phi}) into 
Eq.\ (\ref{eq:rigidrotation}) yields
\bea
\label{eq:lhs}
U(\varphi) &=&
\Omega \; {\bf n}({\bf R})\cdot \left( {\bf e}_z \times {\bf R}\right) \\
&=& - \frac{i \Omega}{| \bm{\tau}(\varphi) |}  
        \left[\sum_{l} l \, b_l \, e^{i l \varphi} 
        + \sum_{l} \sum_{p}
        \frac{l}{2} \, b_{l\!-\!p} b_p \,  e^{i l \varphi} \right] . \nn
\eea
We seek frequencies $\Omega$ and coefficients $b_l$ that satisfy Eqs.\
(\ref{eq:normalv}) and (\ref{eq:lhs}).  The solutions to this 
{\em non-linear eigenvalue problem} represent edge modes that
propagate without any dispersion.

\end{multicols}
\section{Solving the non-linear eigenvalue problem---perturbative approach}
 \vspace{-.3cm}
\label{sec:solving-nep}

Unfortunately, it has not been possible to solve 
exactly the non-linear eigenvalue problem for $b_l$ and
$\Omega$ as written in Eqs.\ (\ref{eq:normalv}) 
and (\ref{eq:lhs}). We therefore seek a perturbative solution by expanding 
the right-hand side of Eq.\ (\ref{eq:normalv}) in powers of $b_l$. This
allows us to to go beyond the linear approximations used in the past,
and we have succeeded in calculating shape deformations to 
${\cal O}[b_l^4]$ and angular frequencies to ${\cal O}[b_l^5]$. 
Expanding the non-linear eigenvalue problem to fifth order, we find
the condition
\begin{eqnarray}
\label{eq:NLEP}
\tilde{\Omega} \left(b_l + \frac{1}{2} 
\sum_p b_{l-p} b_p \right) &=&
Q_l\, b_l 
+ \sum_p R_{l\!-\!p,p} \, b_{l-p} b_p 
+\! \sum_{p,q} S_{l\!-\!p,p\!-\!q,q} \, b_{l-p} b_{p-q} b_q  
\nonumber \\*
&&+ \!\sum_{p,q,r} T_{l\!-\!p,p\!-\!q,q\!-\!r,r} \, b_{l-p} 
        b_{p-q} b_{q-r} b_r 
+\!\!\! \sum_{p,q,r,s} U_{l\!-\!p,p\!-\!q,q\!-\!r,r\!-\!s,s} 
        b_{l-p} b_{p-q} b_{q-r} b_{r-s}
b_s + {\cal O} [b_l^6] \,, 
\end{eqnarray}
%
where 
\be
\label{eq:omega_tilde}
\tilde{\Omega} = \frac{\epsilon m_e \omega_c R_0}{\bar{n} e^2} \,\Omega,
\ee
%
and the ``matrix elements'' $Q$, $R$, $S$, $T$ and $U$ are obtained
from Eq.\ (\ref{eq:normalv}) in Appendix \ref{sec:matrix}. 

Since the equation to be solved results from an expansion in powers of
$b_l$, it is sufficient to solve Eq.\ (\ref{eq:NLEP}) 
perturbatively, by expanding in powers of the largest coefficient
$b_L$. Let us assume that 
\be
{b_l}  \, \left\{
\begin{array}{ll}
= {\cal O}[\delta] \,, &\;\; {\rm for} \;\; l= \pm L \,, \\*
= {\cal O}[\delta^2] \,, &\;\; {\rm for} \;\; l \neq \pm L \,.
\end{array}
\right.
\ee
We then consider an expansion of the coefficients $b_l$ and eigenvalue
$\tilde{\Omega}$ in powers of $\delta$:
\bea
\label{eq:exp_b}
&&{b_l}  = \left\{
\begin{array}{ll}
 b^{(1)}_L \,, &\;\; {\rm for} \;\; l= \pm L \,, \\
 b^{(2)}_l +  b^{(3)}_l + \cdots \,, &\;\; {\rm for} \;\; l \neq \pm L \,.
\end{array} \right. \\
\label{eq:exp_omega}
&&\tilde{\Omega}_L = \tilde{\Omega}_L^{(0)} + \tilde{\Omega}_L^{(1)} + 
        \tilde{\Omega}_L^{(2)} + \cdots \,
\eea
where $b^{(k)}$ and $\tilde{\Omega}^{(k)}$ are of order 
${\cal O}[\delta^k]$. By substituting expressions
(\ref{eq:exp_b}) and (\ref{eq:exp_omega}) into Eq.\ (\ref{eq:NLEP}), and
grouping terms order by order, we find that:

\noindent
\begin{enumerate}

\item
The first-order term yields the dispersion relation in the
linear approximation: 
\be
\label{eq:omega-0}
\tilde{\Omega}^{(0)}[L] = Q_L \,;
\ee

\item
Second-order terms couple the fundamental and second harmonics
of the deformation, with no correction to the eigenvalue: 
\bea
&&\tilde{\Omega}^{(1)}[L] = 0 \,, \nn \\
&&b^{(2)}_{\pm 2L} = \frac{1}{2}\,\frac{2 R_{L,L} - Q_L}{Q_L - Q_{2L}} \,
        [b_L^{(1)}]^2  \,, \\
&&b^{(2)}_{l}  = 0 \,,\;\; {\rm for} \;\; l \neq \pm 2 L \,;\nn
\eea

\item
Third-order terms give the first correction to the
frequency and couple the first and third harmonics: 
\bea
&&\tilde{\Omega}^{(2)}[L] =  3 S_{L,L,-L} [b_L^{(1)}]^2 +
        (2 R_{2L,-L} - Q_L) b_{2L}^{(2)} \,, \nn \\
&&b^{(3)}_{\pm 3L} = \frac{ (2 R_{2L,L} - Q_L) b_{2L}^{(2)} b_L^{(1)}
        + S_{L,L,L} [b_L^{(1)}]^3}{Q_L-Q_{3L}} \,,\\
&&b^{(3)}_{l}  = 0 \,,\;\; {\rm for} \;\; l \neq \pm 3 L \;;\nn
\eea

\item
Fourth-order terms provide coupling to both second and fourth harmonics:
\bea
&&\hspace{-.5cm} \tilde{\Omega}^{(3)}[L] = 0 \,, \nn \\
&&\hspace{-.5cm}  b^{(4)}_{\pm 2L} = \frac{1}{Q_L - Q_{2L}} \left\{
        ( 2 R_{3L,-L}-Q_{L} ) b^{(3)}_{3L} b_L^{(1)} 
   - \tilde{\Omega}^{(2)}[L] \left( b^{(2)}_{2L} +
	\frac{[b_L^{(1)}]^2}{2}\right) 
        + 6 S_{2L,L,-L} b^{(2)}_{2L} [b_L^{(1)}]^2 
        + 4 T_{L,L,L,-L} [b_L^{(1)}]^4 \right\} \!, \nn \\
&&\hspace{-.5cm} b^{(4)}_{\pm 4L} = \frac{1}{Q_L - Q_{4L}} \left\{
        \frac{1}{2} ( 2 R_{2L,2L}-Q_{L} ) [b^{(2)}_{2L}]^2
        + ( 2 R_{3L,L}-Q_{L} ) b^{(3)}_{3L} b_L^{(1)}
        + 3 S_{2L,L,L} b^{(2)}_{2L} [b_L^{(1)}]^2
        + T_{L,L,L,L} [b_L^{(1)}]^4 \right\} \!,\!\!\!  \\
&&\hspace{-.5cm} 
b^{(4)}_{l}  = 0 \,, \;\; {\rm for} \;\; l \neq \pm 4 L \;;\nn
\eea

\item
The fifth-order term follows the same pattern. The
couplings to higher harmonics is quite complex and we chose not to
pursue it. We show only the correction to the eigenvalue:
\bea
\tilde{\Omega}^{(4)}[L] &=& (2 R_{2L,-L} - Q_L) b^{(4)}_{2L} 
  + (2 R_{3L,-2L}-Q_{L}) \frac{b^{(3)}_{3L} b^{(2)}_{2L}}{b^{(1)}_{L}}
  - \tilde{\Omega}^{(2)}[L] b^{(2)}_{2L} 
  + 3 S_{3L,-L,-L} b^{(3)}_{3L} b^{(1)}_{L}  \nn \\*
&&  
  + 6 S_{L,2L,-2L} [b^{(2)}_{2L}]^2
  + (12 T_{2L,L,-L,-L} + 4 T_{-2L,L,L,L} ) b^{(2)}_{2L} [b_L^{(1)}]^2 
  + 10 U_{L,L,L,-L,-L} [b_L^{(1)}]^4 \,.
\eea

\end{enumerate}

\noindent
Note that the largest term in the perturbative expansion corresponds
to the lowest harmonic and that higher order elements preserve the
rotational symmetry $C_L$ (rotations by $2 \pi/L$) of the initial
term.

\begin{multicols}{2}
\section{Solutions for the two-dimensional electron system}  \vspace{-.3cm}
\label{sec:solving-nep-2DES}

We now show some invariant shapes for the 2DES (see
Secs. \ref{sec:dyn} and \ref{sec:nep}).  Coefficients $b_l$ and
frequencies $\tilde{\Omega}$ are obtained using the {formul\ae} of
Sec.\ \ref{sec:solving-nep} and matrix elements $Q$, $R$, $S$, $T$ and
$U$ calculated in Appendix \ref{sec:matrix}.  

Table \ref{table:2des} summarizes these results for states of
rotational symmetry $C_L$, with $L$ = 2, 3, 4 and 5.  
Some particular shapes obtained from these coefficients and
Eq.\ (\ref{eq:paramet}) are shown in Fig. \ref{fig:shapes} [for a
comparison with similar states for vortex patches see Figs.\
\ref{fig:kirchhoff} and \ref{fig:dz}].

For large deformations, the appearance of oscillations indicate that
higher order terms are needed, since the perturbative method
corresponds to a truncation of the Fourier decomposition
[Eq.\ (\ref{eq:paramet})].  An alternative approach, based on a local
induction approximation, provides a better description in those cases.
This alternative formulation is presented in Sec.\ \ref{sec:lia}.

It is interesting to note that the linear result for the frequency is
[see Eqs.\ (\ref{eq:omega-0}) and (\ref{eq:ql})]
\begin{equation}
\label{eq:Q}
\tilde{\Omega}^{(0)}[l]  =  Q_l =
4  \sum_{k=2}^{|l|}\frac{1}{2k-1} \,,
\end{equation}
where the last sum can be related to the digamma function \cite{as}
$\psi(|l|\!+\!{\small 1/2})$. This linear result has been
previously derived by several authors;\cite{linear} corrections are
${\cal O}[b_l^2]$. For a direct comparison with Eq.\
(\ref{eq:omega0}), which corresponds to the {\em large-$l$} limit, we
substitute the asymptotic expansion for the sum in Eq.\ (\ref{eq:Q});
multiplication by $R_0$ yields the propagation velocity: 
\begin{equation}
v_g = -2 \ln \left( \frac{e^{-\gamma}}{4 \, e^2 |l|} \right)
        \frac{\bar{n} e^2}{\epsilon m_e \omega_c} \,,
\end{equation}
which closely corresponds to the group velocity
$v_g \!\equiv\! \partial \omega_0(k) / \partial k$ obtained from Eq.\
(\ref{eq:omega0}) after the substitution $l\!\sim\!ka$.  The
dispersion for these linear edge excitations has been confirmed
experimentally in both the frequency \cite{FreqDomain} and time
\cite{AcousticModes,TimeDomain,TimeDomain2} domains. 

\end{multicols}
\begin{table}
\caption{ \label{table:2des}
  Stationary deformations of a two-dimensional electron system. 
  Angular frequencies and lowest harmonics of the deformation
  for $L\!=\!2,3,4,5$.} 
\begin{center}
\begin{tabular}{|c||c|c|c|c|}
\hline 
&&&&\\
$\;\;l\;\;$ & ${\tilde{\Omega}=({\epsilon m_e \omega_c R_0}/{\bar{n}
        e^2})\, \Omega}$ 
        & $b_{2l}$ & $b_{3l}$ & $b_{4l}$\\
&&&&\\
\hline \hline
&&&&\\
2 & $\frac{4}{3} - 5.09048 \, b_2^2 - 16.1333 \, b_2^4$ 
        & $\frac{29}{24}\, b_2^2 + 1.86911 \, b_2^4$ 
        & $0.911769 \, b_2^3$ & $-0.345388 \, b_2^4$ \\ 
&&&&\\
3 & $\frac{32}{5} - 16.8464 \, b_3^2 + 39.6332 \, b_3^4$
        & $1.63473 \, b_3^2 + 8.36452 \, b_3^4$ 
        & $1.97442 \, b_3^3$ & $-0.540557 \, b_3^4$ \\ 
&&&&\\
4 & $\frac{284}{105} - 30.0989 \, b_4^2 - 154.821 \, b_4^4$ 
        & $1.94074 \, b_4^2 + 24.3045 \, b_4^4$ 
        & $2.56102 \, b_4^3$ & $-3.45272 \, b_4^4$ \\ 
&&&&\\
5 & $\;\frac{992}{315} - 45.5507 \, b_5^2 - 418.524 \, b_5^4\;$
        & $\;2.17945 \, b_5^2 + 50.1748 \, b_5^4\;$ 
        & $\; 2.74603 \, b_5^3$ & $ -9.80691 \, b_5^4\;$ \\
&&&&\\
\hline
\end{tabular}
\end{center} 
\end{table}

\begin{multicols}{2}
\section{Local induction approximation}  \vspace{-.3cm}
\label{sec:lia}

As we have seen, 
the motion of the edge is determined by the velocity of the fluid at
the surface. The nonlocal equation for the velocity of the boundary,
Eq.\ (\ref{eq:vlineint}), can be turned into a differential equation
for the {\em curvature} of the boundary if we concentrate on the local
contributions. This {\em local induction approximation} (LIA)
\cite{batchelor} was explored by Goldstein and Petrich in a 
series of papers dealing with the evolution of vortex
patches.\cite{VortexPatchDynamics,KdVHierarchy}  The situation is
considerably more favorable in this problem, due to the more rapid 
decay of the  interaction [$1/r$ for charges vs.\ $\ln(r)$ for
vortices, see Eqs.\ (\ref{eq:vlineint}) and (\ref{eq:vp})].  Figure
\ref{fig:droplet} defines most terms used in this section.

The LIA corresponds to the introduction of a large distance cut-off
$\Lambda$ in the expression for the velocity of the boundary 
${\bf v}[{\bf r}(s)]$ [Eq.\ (\ref{eq:vlineint})]:
\be
\label{eq:truncation}
\oint_\Gamma \{\cdots\} \, ds' \; \bm{\longrightarrow} \;
\int_{s-\Lambda/2}^{s+\Lambda/2} \!\!\!\! \{\cdots\} \, ds' \,.
\ee

The line integral in Eq.\ (\ref{eq:vlineint}) is then calculated by
expanding the integrand in powers of $\Delta \!\equiv\! (s'\!-\!s)$.
By using the Frenet-Serret relations 
\be
\label{eq:fs}
\begin{array}{l}
        {\bf r}_s = {\bf t} \,, \\
        {\bf t}_s = -\kappa \, {\bf n} \,,\\
        {\bf n}_s = \kappa \, {\bf t} \,,
        \end{array} 
\ee
where $\kappa\!=\!\theta_s$ is the local curvature of the boundary, 
we obtain
\bea
\label{eq:lia1}
{\bf t}(s') & \simeq & {\bf t}(s) \left[ 1 -
        \frac{\Delta^2}{2}\kappa^2 - \frac{\Delta^3}{2} \kappa \kappa_s 
        + ... \right]  \\*
&&        + {\bf n}(s) \left[ - \Delta \, \kappa 
        - \frac{\Delta^2}{2}\kappa_s 
        + \frac{\Delta^3}{6}(\kappa^3 - \kappa_{ss})
        + ... \right] \,,\nn  
\eea
\bea
\label{eq:lia2}
{\bf r}(s')&-&{\bf r}(s) \simeq
        {\bf t}(s) \left[
        \Delta - \frac{\Delta^3}{6} \kappa^2 
        - \frac{\Delta^4}{8}\kappa \kappa_s 
        + ... \right]  \\
&&        + {\bf n}(s) \left[ - \frac{\Delta^2}{2}\kappa
        - \frac{\Delta^3}{6} \kappa_s 
        + \frac{\Delta^4}{24} ( \kappa^3 - \kappa_{ss} )
        + ... \right] \,. \nn  
\eea

To lowest order the normal and tangential velocities then are given by:
\begin{eqnarray} 
\label{eq:LIAresults}
&&U_{\rm LIA} = - \left[\frac{\bar{n} e^2}{\epsilon m_e \omega_c}\right]
         \frac{\Lambda^2}{8} \, \kappa_s \,, \nonumber \\
&&W_{\rm LIA} = \left[\frac{\bar{n} e^2}{\epsilon m_e \omega_c}\right]
         \left( \ln \frac{\Lambda^2}{2 r_0} - \frac{11 \Lambda^2}{96}
           \, \kappa^2 \right) .
\end{eqnarray}
It is worth noting that since the rate of change of the area ${\cal A}$
of the droplet is  ${\cal A}_t \!=\! \oint ds\, U(s)$ [see 
Eq.\ (\ref{eq:At})], the LIA with $U_{\rm LIA} \!\propto\! \kappa_s$
(or any exact differential) automatically conserves area.  This is not
surprising since we started with an incompressible system, but shows
that the local approximation used has not introduced an obvious error.
It is also interesting to realize that the perimeter ${\cal L}$ of the
curve derived from these velocities is conserved as well:  
${\cal L}_t \!=\! \oint ds \, (\kappa U + W_s) = 0$ 
[see Eq.\ (\ref{eq:Lt})].  

The time evolution of a curve in two dimensions is given quite
generally by the integro-differential equation
\cite{KdVHierarchy,Brower} (see Appendix \ref{sec:curves})
\begin{equation}
\label{eq:diffKappa}
\kappa_t = - [ \kappa^2+\partial_{ss} ] U + \kappa_s W - \kappa_s
\int_0^s [\kappa U + W_{s'}]ds' \,.
\end{equation}
We now introduce the results from Eq.\ (\ref{eq:LIAresults}).  A 
``gauge'' change, realized by modifying $W$ so that the integrand of
Eq.\ (\ref{eq:diffKappa}) vanishes, eliminates the remaining non-local
dependences and yields 
\be
\label{eq:diffKappa_gauged}
\kappa_t = \left[\frac{\bar{n} e^2}{\epsilon m_e \omega_c}\right]
         \frac{\Lambda^2}{8} \left( 
	\frac{3}{2} \kappa^2 \kappa_s + \kappa_{sss}  \right) .
\ee
After a simple time rescaling, the curvature satisfies the mKdV
equation:\cite{mKdV} 
\begin{equation}
\label{eq:mkdv}
\kappa_t = 
\frac{3}{2} \kappa^2 \kappa_s +
  \kappa_{sss}  \,.
\end{equation}
The mKdV dynamics are integrable, with an infinite number of globally
conserved geometric quantities,\cite{VortexPatchDynamics}  the most
important of which are the center of mass, area, and angular
momentum of the droplet.  

\begin{figure}
\begin{center}
\leavevmode
\epsfbox{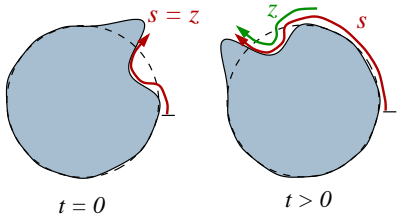}
\end{center}
\caption{\label{fig:travel}
  A uniformly propagating edge deformation.  The curvature satisfies
  $\kappa(s,t) = \kappa(z)$, with $z=s-c\, t$. }
\end{figure}

The mKdV equation possesses a variety of soliton solutions, 
including traveling wave solutions and propagating ``breather''
solitons (see Appendix \ref{sec:infinite}).  Here we will focus on the
traveling wave solutions of Eq.\ (\ref{eq:mkdv}) of the form
\be
\label{eq:travel}
\kappa(s,t) = \kappa(z) \,, \;\;\;\; {\rm with} \;\; 
        z \equiv s - c \,t \,,
\ee
which represents uniformly rotating deformed droplets (see
Fig. \ref{fig:travel}). The ordinary differential equation for
$\kappa(z)$ can be integrated twice with the result  
\begin{equation}
\label{eq:ode}
\frac{1}{2} (\kappa')^2 = -\frac{1}{8} \kappa^4 +
        \frac{1}{2} c \kappa^2+ a \kappa - 2 b
\,,
\end{equation}
where $a$ and $b$ are constants of integration ($a\!=\!b\!=\!0$ for
infinite systems, see Appendix \ref{sec:infinite}).  This ordinary
differential equation can be easily integrated: 
\bea
\label{eq:integral}
&& z - z_0 = \pm \int_{\kappa_0}^{\kappa} 
        \frac{d \kappa'}{\sqrt{-V(\kappa')}} \,, \\
\label{eq:Vg}
&& V(\kappa) = \frac{\kappa^4}{4} - c \, \kappa^2 - 
        2 a \, \kappa + 4 b \,.
\eea

\begin{figure}
\begin{center}
\leavevmode
\epsfbox{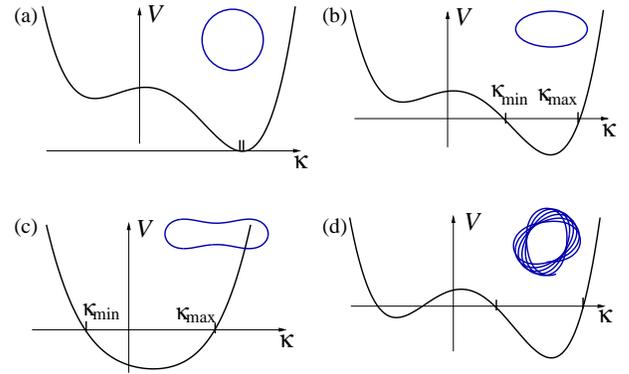}
\end{center}
\caption{\label{fig:potential}
        Some possible potentials $V(\kappa)$ and corresponding
        schematic solutions for the shape of the boundary. Potentials
        with two real zeros may have 
        physical solutions: (a) Curvature is fixed, the solution is a
        circle; (b) Curvature is positive definite, the boundary is
        convex; (c) Positive and negative curvatures.  (c) Potentials
        with four zeros have solutions with unphysical
        self-intersections.}
\end{figure}

This problem is thus analogous to a particle moving in a quartic
potential (Fig. \ref{fig:potential}). The integrals involved can be
expressed in terms of elliptic integrals \cite{as} and depend
crucially on the nature and location of the zeros of $V(\kappa)$. For
curves with finite perimeter and no self-crossings we find that it is
necessary to have two real and two complex zeros: $\kappa_{\rm max}$,
$\kappa_{\rm min}$, and  $-(\kappa_{\rm max} \!+\! \kappa_{\rm min})/2
\!\pm\! i \xi$ (see Appendix \ref{sec:elliptic}).  Note that 
$\kappa_{\rm max}$ and $\kappa_{\rm min}$ correspond to the maximum
and minimum curvatures of the boundary.

\begin{figure}
\begin{center}
\leavevmode
\epsfbox{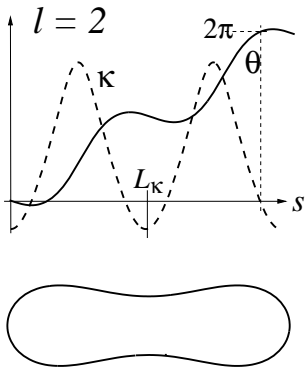}
\end{center}
\caption{\label{fig:pbc}
        Illustration of the boundary conditions implied by
        Eq.\ (\protect\ref{eq:periodic}).  Top: after $l$ periods
        $L_\kappa$ of the curvature $\kappa(s)$, the tangent angle
        $\theta$ increases by precisely $2 \pi$. Bottom: the resulting
        deformed boundary.}
\end{figure}

The periodic solutions of Eq.\ (\ref{eq:integral}), expressed
in terms of Jacobi elliptic functions,\cite{as} are given by 
(Appendix \ref{sec:elliptic}):
\be
\label{eq:curvatureMKDV}
\kappa(z) = \frac{(q \kappa_{\rm max} \!+\! p \kappa_{\rm min}) + 
        (p \kappa_{\rm min} \!-\! q \kappa_{\rm max})\,
  {\rm cn}\! \left( \!\sqrt{p q} \frac{\displaystyle z}{ 2}\,
        |\, \lambda \right) }
{(p+q)+(p-q)\,{\rm cn}\! \left( \!\sqrt{p q} 
\frac{\displaystyle z}{ 2}\,|\,\lambda \right)} , 
\ee
where
\bea
p &=& \sqrt{(3 \kappa_{\rm max} + \kappa_{\rm min})^2 +
\xi^2} \,,\nonumber \\ 
q &=&  \sqrt{(\kappa_{\rm max} + 3\kappa_{\rm min})^2 +
\xi^2} \,, \\ 
\lambda &=& 
        \sqrt{ \left[(\kappa_{\rm max} - \kappa_{\rm min})^2 -
    (p-q)^2\right]/4 p q} \,. \nn
\end{eqnarray}
The free parameter $\xi$ is actually determined by the {\em boundary
conditions}. The period of $\kappa$ is given by the elliptic integral 
\be
\label{eq:Lk}
L_\kappa = \frac{8}{\sqrt{p q}} \, K(\lambda) \,.
\ee
We now require that the tangent angle increases by a factor of $2 \pi$
after an integer multiple $l$ curvature periods, so that the curve is
closed and with no self crossings (see Fig. \ref{fig:pbc}):
\be
\label{eq:periodic}
\theta(l\,\! L_\kappa) = \int_0^{l\,\! L_\kappa} 
        \!\!\!\kappa(s) \, ds = 2 \pi \,.
\ee
It is evident
that the resulting curves are invariant under rotations by $2 \pi/l$,
that is the curves have $C_l$ symmetry. The curves thus generated can
be characterized by $(l,\kappa_{\rm max},\kappa_{\rm min})$ or more
conveniently, although indirectly, by the symmetry, the area and the
perimeter of the curve. 

The contour shape can be easily determined once the tangent angle
$\theta(s)$ is known as a function of arc-length.  Since 
$ d z \equiv dx + i \, dy = \exp[{i \, \theta(s)}]  \, ds $, 
we have 
\be
\label{eq:shape-int}
z(s) \equiv x(s) + i \, y(s) = \int_0^s e^{i \, \theta(s')}  \, ds' \,.
\ee

\begin{figure}
  \begin{center}
    \leavevmode
    \epsfxsize=4.5in
    \epsffile{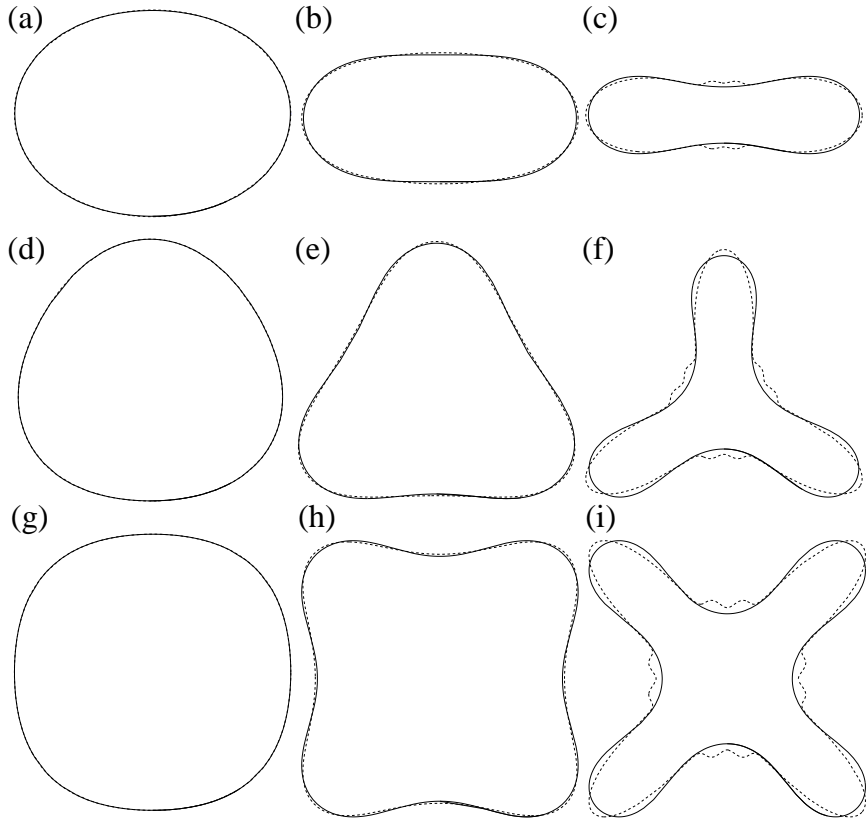}
  \end{center}
  \caption{ \label{fig:shapes} 
        Uniformly rotating shapes of a 2DES. Solid lines: solutions of
the mKdV equation obtained from the local induction
approximation. Dotted lines: solutions obtained using the perturbative
expansion. The values of the coefficient $b_l$, and the ratio of
curvatures $\sigma\!\equiv\!\kappa_{\rm min}/\kappa_{\rm max}$ are:
{(a)} $b_2 \!=\! 0.073, \, \sigma \!=\! 0.4$;
{(b)} $b_2 \!=\! 0.19, \, \sigma \!=\! 0$;
{(c)} $b_2 \!=\! 0.36, \, \sigma \!=\! -\!0.2$;
{(d)} $b_3 \!=\! 0.027, \, \sigma \!=\! 0.4$;
{(e)} $b_3 \!=\! 0.10, \, \sigma \!=\! -\!0.2$;
{(f)} $b_3 \!=\! 0.29, \, \sigma \!=\! -\!0.45$;
{(g)} $b_4 \!=\! 0.014, \, \sigma \!=\! 0.4$;
{(h)} $b_4 \!=\! 0.089, \, \sigma \!=\! -\!0.4$;
{(i)} $b_4 \!=\! 0.24, \, \sigma \!=\! -\!0.56$. }
\end{figure}

The full lines in Fig.\ \ref{fig:shapes} show some uniformly rotating
soliton shapes,  calculated from Eqs.\ (\ref{eq:curvatureMKDV}) and 
({\ref{eq:shape-int}}). These are essentially identical  with
Goldstein and Petrich's \cite{KdVHierarchy} soliton solutions for the
vortex patch problem, but the more local nature of the interaction in
the 2DES case should guarantee a better correspondence with the exact
solutions including non-local terms. Indeed, the curves resulting from
the perturbative method and the LIA are quite close, even for
considerable deformations of the boundary. For larger deformations the
perturbative results show artifacts due to the limited number of
Fourier components. The advantage of the LIA becomes evident in this
case, since it is an expansion in powers of the  {\em curvature} and
not the  {\it deformation}, and thus allows for relatively large
long-wavelength deformations. 
More significantly, the 
LIA and the resulting integrable dynamics allow one to uncover
geometrical conservation laws which would be hidden in a perturbative
calculation.\cite{conservation}
This advantage comes at a price: the detailed information on
frequencies is obscured by the introduction of the long distance
cut-off $\Lambda$ and by the gauge transformation of the tangential
velocity $W$, while the frequency is easily obtained in the
perturbative calculation.

\end{multicols}

\begin{multicols}{2}

\section{Conclusions}  \vspace{-.3cm}

A contour dynamics formulation of the
excitations on the edge of a two-dimensional electron system in a
magnetic field has allowed us to demonstrate the existence, beyond
the usual linear regime, of shape deformations that propagate
uniformly.  A local approximation to the nonlocal dynamics shows that 
the curvature of the edge of the droplet obeys the modified 
Korteweg-de Vries equation, which has integrable dynamics and soliton
solutions.  Earlier studies \cite{zhitenev} of edge channels in QH
samples have shown the presence of nonlinear waves, but in Ref.\
\onlinecite{zhitenev} the nonlinearity originates in the variations of
the strength of the confining field (the ${\bf E}_{\rm ext}$ of Sec.\
\ref{sec:hydro2DES}), whereas here we concentrate on nonlinear effects
originating in geometrical effects.

\begin{figure}
\begin{center}
\leavevmode
\epsfbox{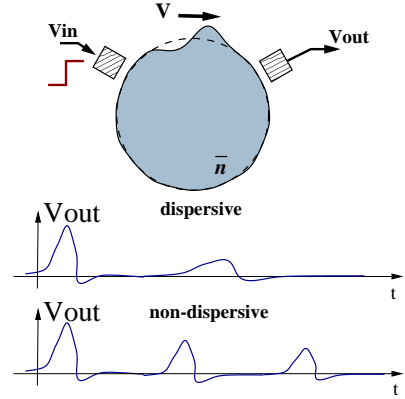}
\end{center}
\caption{\label{fig:dispersive}
        Soliton excitations travel without shape changes.  In a
        circular QH system\cite{TimeDomain} with non-contact probes
	it is possible to observe repeated signals.  While a gradual
	decrease of the amplitude is always expected due to residual
	dissipation, dispersive and non-dispersive modes may be
	distinguished by the preservation of the general shape and
	width of the non-dispersive modes.} 
\end{figure}

Since these solutions are dispersionless, it may be possible to
distinguish them from linear edge waves in time-of-flight
measurements of the type depicted in Fig.\ \ref{fig:dispersive}.
In a circular QH system\cite{TimeDomain} a voltage pulse applied to a
gate produces an edge deformation.  The deformation propagates along
the edge of the system in one direction and is detected, i.e. by means
of a capacitive probe.  For this geometry, the pulse comes back
repeatedly and it is possible to observe it after numerous passes.
While a gradual decrease of the amplitude is always expected due to
residual dissipation, dispersive and non-dispersive modes may be
distinguished by the preservation of the general shape and width of
the non-dispersive modes. 

On the theoretical side, it would be interesting to
connect our hydrodynamic treatment of these edge solitons with 
field-theoretical treatments of edge excitations.\cite{linear,winf}

\acknowledgments \vspace{-.3cm}
We would like to thank Raymond Goldstein for useful discussions.
This work was supported by the NSF grant DMR-9628926.


\end{multicols}

\medskip
\appendix
\noindent \hrule

\section{Matrix elements for the non-linear eigenvalue problem}  \vspace{-.3cm}
\label{sec:matrix} 

In order to determine the matrix elements $Q$, $R$, $S$, $T$ and $U$
for the non-linear eigenvalue problem [Eq.\ (\ref{eq:NLEP})], we need to 
expand Eq.\ (\ref{eq:normalv}) in powers of the coefficients $b_l$ of
the parametrization [Eq.\ (\ref{eq:paramet})]. We initially realize
that 
\bea
{\bf n}(\varphi) {\bf \cdot} \bm{\tau}(\varphi + \omega) &=&
        \frac{1}{| \bm{\tau}(\varphi) |} \Biggl\{
        - \sin \omega - 2 \sum_{l} b_l   \, e^{i l \varphi}
         \left[ \cos \frac{l \omega}{2} \, \sin \omega + 
                l \cos \omega \, \sin \frac{l \omega}{2}\right] +
\nn \\*
&&  
   + \sum_{l,p} b_{l\!-\!p} b_p \, 
        e^{i l \varphi} \left[ 
        (lp \!-\! p^2 \!-\! 1) \, \cos\frac{(l\!-\!2p)\omega}{2} \,
        \sin\omega \!-\! 
        (l \!-\! 2p)  \, \cos \omega \, 
        \sin \frac{(l\!-\!2p)\omega}{2} \right]
 \Biggr\} , 
\eea
\bea
|{\bf R}(\varphi+\omega) - {\bf R}(\varphi)|^2 &=& \nn\\*
        && \hspace{-3cm} 
        4 \sin^2 \frac{\omega}{2} \left[ 1 + 2 \sum_l b_l \, 
        e^{i l \omega/2} \, \cos \frac{\omega}{2}  
        - \sum_{l,p}  b_{l\!-\!p} b_p \, e^{i l \omega/2} \,
        \frac{\sin \frac{(l-p+1)\omega}{2} \, \sin \frac{(p-1)\omega}{2} +
        \sin \frac{(l-p-1)\omega}{2}\, \sin \frac{(p+1)\omega}{2} }
        {2 \sin^2 \frac{\omega}{2}} \right] .
\eea
By expanding ${\bf n}(\varphi) {\bf \cdot} \bm{\tau}(\varphi + \omega) /
|{\bf R}(\varphi+\omega) - {\bf R}(\varphi)|$ in powers of $b_l$, and
integrating over $\omega$ [see Eq.\ (\ref{eq:normalv})], one obtains
the relevant matrix elements $Q$, $R$, $S$, $T$ and $U$.  All
integrals converge without the need for short distance cut-offs.  The
three lowest order terms can be written as follows:
\bea
\label{eq:ql}
Q_l &=& 4  \sum_{k=2}^{|l|}\frac{1}{2k-1} \,,\\
\label{eq:rl}
R_{l-p,p} &=& \sum_{k=1}^{|l|}\frac{1}{2k-1} 
        - \sum_{k=1}^{|l-p|}\frac{1}{2k-1} 
        - \sum_{k=1}^{|p|}\frac{1}{2k-1} \,, \\
\label{eq:sl}
S_{l\!-\!p,p\!-\!q,q} &=& - \frac{5}{l} \left[ 
       {\frac{l}{1 - 4{l^2}}} + 
       {\frac{p}{1 - 4{p^2}}} + 
       {\frac{q}{1 - 4{q^2}}} + 
       {\frac{l - p}{1 - 4{{( l - p ) }^2}}}  + 
       {\frac{l - q}{1 - 4{{( l - q ) }^2}}}  + 
       {\frac{p - q}{1 - 4{{( p - q ) }^2}}}  + 
       {\frac{l - p + q}{1 - 4{{( l - p + q ) }^2}}} 
       \right] \nn \\
       &&+
\frac{1}{12} \left[ 
-(3 + 4l^2) \sum_{k=1}^{|l|}\frac{1}{2k-1} - 
(1 + 4p^2)\sum_{k=1}^{|p|}\frac{1}{2k-1} + 
(5 + 4q^2)\sum_{k=1}^{|q|}\frac{1}{2k-1} + 
[5 + 4(l - p)^2] \sum_{k=1}^{|l-p|}\frac{1}{2k-1} \right. \nn \\
&&\left. - 
[1 + 4(l - q)^2]\sum_{k=1}^{|l-q|}\frac{1}{2k-1}+ 
[5 + 4(p - q)^2]\sum_{k=1}^{|p-q|}\frac{1}{2k-1} - 
[1 + 4(l - p + q)^2]\sum_{k=1}^{|l-p+q|}\frac{1}{2k-1} \right] \,.
\eea
Higher order terms are long and uninspiring.




\begin{multicols}{2}
\section{Comparison with the vortex patch case}  \vspace{-.3cm}
\label{sec:vortex-patch}

For the sake of comparison, we draw analogy to the case of a vortex 
patch, a two-dimensional, bounded region of constant vorticity
$\omega_p$ surrounded by an irrotational fluid.\cite{lamb} 
The vorticity can either be distributed, as in a regular fluid, or
concentrated in individual vortices, in the case of a superfluid
(in this case it is clear that the hydrodynamic treatment will be
valid only for length-scales larger than the inter-vortex spacing).
The important thing is that in ideal fluids the area of the vortex
patch is conserved due to  Kelvin's circulation
theorem,\cite{batchelor} and therefore the vortex patch is essentially
incompressible.  Figure \ref{fig:droplet} can be used to describe this
case by replacing the electron density $n$ by the vorticity $\omega_p$. 

The steady state solution in this case is clearly a circle, and small
deformations of the boundary travel along the boundary itself, as has
been known for a long time.\cite{lamb} 
One can also ask what happens when the deformations are large:
are there modes that do not change shape, i.e., solitons?  One
such solution has been known since Kirchhoff's time:  an ellipse with
constant vorticity will rotate uniformly in an ideal fluid.  Numerical
calculations by Deem and Zabusky \cite{VortexPatchDynamics,VStates} in
the 70's obtained additional invariant shapes.  

For inviscid incompressible fluids, the equations of motion for the
fluid are simply given by:
\bea
&& \grad \cdot {\bf v} = 0  \,,\\
&& \grad \times {\bf v} = \omega \,,
\eea
where $\omega$ is the vorticity.  Assuming that the velocity far from
the patch vanishes, the velocity of the fluid in presence of a region
of finite vorticity $\omega_p$ can be expressed as
\begin{equation}
\label{eq:vp}
{\bf v_v}({\bf r}) = - \frac{\omega_p}{2 \pi} 
        \oint_\Gamma ds' \, {\bf t}(s') \,
        \ln \left[{|{\bf r}-{\bf r}(s')|\over r_0} \right].
\end{equation}

The arguments presented in Secs. \ref{sec:dyn} and \ref{sec:nep} are
now applied {\em mutatis mutandis} to this case. The only difference
with the 2DES comes from the fact that the kernel in the interaction
is now logarithmic, and  Eq.\ (\ref{eq:vlineint}) is then replaced by
Eq.\ (\ref{eq:vp}).  We see that, as in the 2DES case
(Secs. \ref{sec:dyn} and \ref{sec:nep}), the dynamics is chiral, being
determined by the tangent vector; and since the fluid contained within
$\Gamma$ is incompressible the area is conserved. 
The normal velocity of the contour is then given by
\begin{equation}
\label{eq:normalv-vp}
U_{\rm v} (\varphi) = - \frac{\omega_p}{2 \pi} 
        \int_0^{2 \pi} \!\! d\varphi' \,
       {\bf n}(\varphi) \cdot \bm{\tau}(\varphi') 
       \ln \left[\frac{|{\bf R}(\varphi) - {\bf R}(\varphi')|}{r_0}
        \right] \!.
\end{equation}

As before, we seek solutions that satisfy Eqs.\
(\ref{eq:rigidrotation}) [or (\ref{eq:lhs})] and
(\ref{eq:normalv-vp}).  First, we follow the procedure of Appendix
\ref{sec:matrix} to determine the matrix elements $Q$, $R$, $S$, $T$
and $U$.  The first few of these are given by
\bea
\label{eq:matrices_vp-q}
Q^{V}_l &=& \frac{1}{2} - \frac{1}{2 |l|} \,, \\
\label{eq:matrices_vp-r}
R^{V}_{l,p} &=& -\frac{1}{4 |l|} \,. \\
\label{eq:matrices_vp-s}
S^{V}_{l\!-\!p,p\!-\!q,q} &=& \frac{1}{12} \Bigl[ 1 - |l| - |l - p| + 
       |2l - p| + |l - q| \nn \\
&&  - |2l - q| - |p - q| + |l + p - q| \nn \\
&& +\delta (2l - p) + \delta (l - q) + \delta (l + p - q) \Bigr] \,.
\eea

\begin{figure}[h]
  \begin{center}
    \leavevmode
    \epsffile{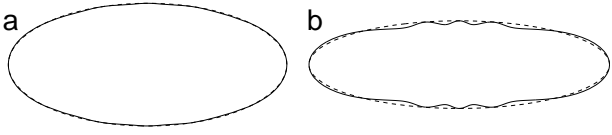}
  \end{center}
  \caption{\label{fig:kirchhoff}
        The $l\!=\!2$ stationary deformations of a vortex patch. Dotted 
        line: the Kirchoff ellipse. Full line: perturbative solution.
        (a) $b_2\!=\!0.2$, (b) $b_2\!=\!0.3$.}
\end{figure}

\begin{figure}[h]
  \begin{center}
    \leavevmode
    \epsffile{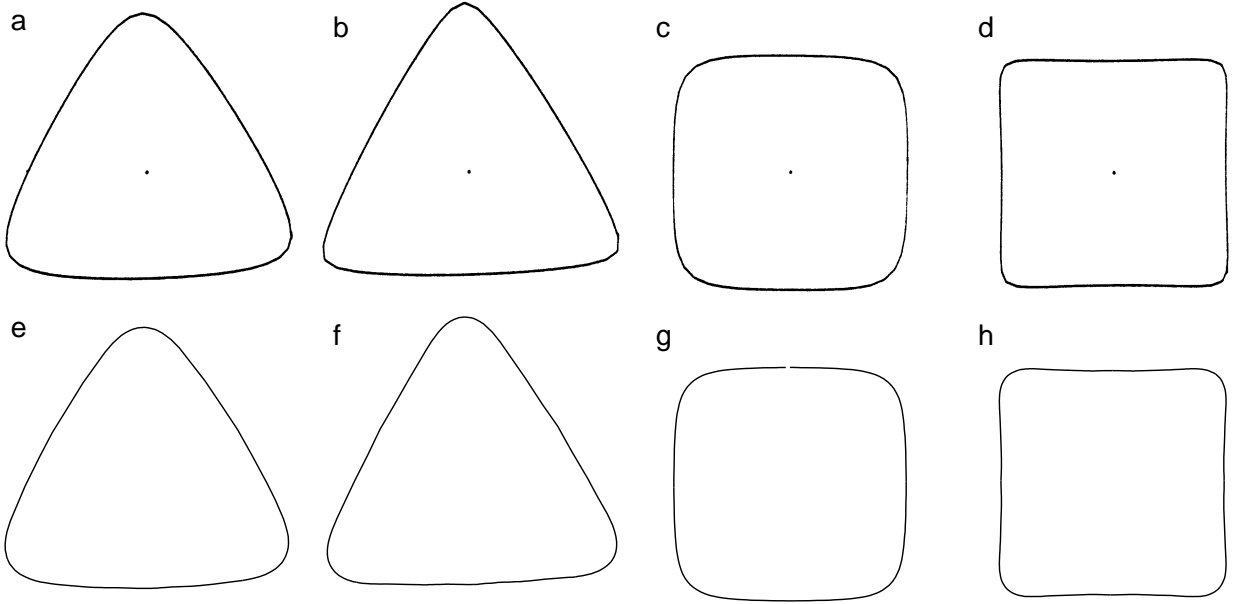}
  \end{center}
  \caption{ \label{fig:dz}
        Vortex-patch ``eigenstates'' with $L\!=\!3$ and $L\!=\!4$. 
        (a)-(d) Obtained numerically by Deem and Zabusky [figures are
        scanned from Ref.\ \protect\onlinecite{VStates}]; 
        (e)-(h) Determined perturbatively.  The lowest harmonic $b_L$
        is taken in each case from Table I of
        Ref.\ \protect\onlinecite{VStates}: $b_3 = 0.096$, 
        $b_3 = 0.11$, $b_4 = 0.048$ and  $b_4 = 0.070$ respectively.
        The solutions are essentially identical and are too close to
        compare effectively in the same picture.}
\end{figure}

We then apply the results of Sec.\ \ref{sec:solving-nep}, to determine
the amplitude of the lowest harmonics for the vortex-patch case, using
the appropriate matrix elements
[Eqs.\ (\ref{eq:matrices_vp-q})--(\ref{eq:matrices_vp-s})].  These 
results are summarized in Table \ref{table:vp}, and some of the
resulting invariant shapes are shown in Figs. \ref{fig:kirchhoff} and
\ref{fig:dz}.

It has long been known \cite{lamb} that an elliptical region of
vorticity in an otherwise irrotational fluid, namely the Kirchhoff
ellipse, rotates uniformly with angular frequency 
$\Omega\!=\!(\omega_p/4) (1\!-\!b^2/a^2)$, where $(a\!+\!b)$ and
$(a\!-\!b)$ represent the maximum and minimum radii respectively 
(see Appendix \ref{sec:kirchhoff}). A simple analysis shows that both
frequency and angular components $b_l$ shown in Table \ref{table:vp}
exactly match a series expansion of the ellipse. Figure
\ref{fig:kirchhoff} compares the perturbative results with the exact
solution. Even for relatively high deformations both results are in
reasonable agreement.

In the case of deformations with higher angular dependences there are
no analytic solutions beyond the linear approximation. In this case, the
angular frequencies are given\cite{lamb} to zeroth-order in the deformation 
by $\Omega\!=\!(m\!-\!1)/2m$, which coincides with the values for
$Q^{V}_l$ [see Eqs.\ \ref{eq:omega-0} and \ref{eq:matrices_vp-q}]. 
Numerical solutions by Deem and Zabusky \cite{VStates} for $l\!=\!3,4$
also compare favorably with the perturbative results, as can be seen in
Fig.\ \ref{fig:dz}. In fact, the angular velocities determined from Table II 
coincide with those shown in Ref.\ \onlinecite{VStates} to all significant
figures of that paper.

\end{multicols}

\begin{table}[h]
\caption{ \label{table:vp}
  Stationary deformations of a vortex patch. Angular frequencies
  and lowest harmonics of the deformation for $L\!=\!2,3,4,5$ [the
  equivalent results for the 2DES are summarized in Table \ref{table:2des}].}
\begin{center}
\begin{tabular}{|c||c|c|c|c|}
\hline 
&&&&\\
$\;\;l\;\;$ & ${\tilde{\Omega}=\Omega/\omega_p}$ 
        & $b_{2l}$ & $b_{3l}$ & $b_{4l}$\\
&&&&\\
\hline \hline
&&&&\\
2 & $\frac{1}{4} - b_2^2 + b_2^4$ 
        & $\frac{3}{2}b_2^2 - \frac{1}{2}b_2^4$ 
        & $\frac{5}{2}b_2^3$ & $\frac{35}{8} b_2^4$ \\ 
&&&&\\
3 & $\frac{1}{3} - 2 b_3^2 - 20 b_3^4$
        & $\frac{5}{2}b_3^2+16b_3^4$ 
        & $8b_3^3$ & $\frac{231}{8}b_3^4$ \\ 
&&&&\\
4 & $\frac{3}{8}-3b_4^2-57b_4^4$ 
        & $\frac{7}{2}b_4^2+\frac{135}{2}b_4^4$ 
        & $\frac{33}{2}b_4^3$ & $\frac{715}{2}b_4^4$ \\ 
&&&&\\
5 & $\;\frac{2}{5}-4b_5^2-152b_5^4\;$
        & $\;\frac{9}{5}b_5^2+172b_5^4\;$ 
        & $\;28b_5^3$ & $\frac{1615}{8}b_5^4\;$ \\
&&&&\\
\hline
\end{tabular}
\end{center} 
\end{table}

\begin{multicols}{2}
\section{The Kirchhoff ellipse as a solution of a contour dynamics
	problem}  \vspace{-.3cm}
\label{sec:kirchhoff}

It is illustrative to show, starting from the contour dynamics, that
an ellipse is indeed an invariant deformation for a vortex patch.
Additionally it can be shown that this is an inherent property of the
logarithmic kernel [Eq.\ \ref{eq:vp}] and that an ellipse it is 
{\em not} a solution for the 2DES.

Consider the following parametrization of an ellipse:
\be
       \begin{array}{l}
                x = (a+b) \cos \eta \,, \\
                y = (a-b) \sin \eta \,.
        \end{array} 
\ee
It is evident that the radius is given by 
\be
\label{eq:radius_ke}
r(\eta) = \sqrt{a^2+2ab \cos 2\eta + b^2} \, ,
\ee
and that tangential and normal vectors are given by
\bea
\bm{\tau} &\equiv&\frac{\partial {\bf r}}{\partial \eta} = 
    - (a+b) \sin \eta \, {\bf e}_x + (a-b) \cos \eta \, {\bf e}_y \,, \\
{\bf n} &=& -\frac{{\bf e}_z \times \bm{\tau}}{|\bm{\tau}|} =
    \frac{ (a-b) \cos \eta \, {\bf e}_x + (a+b) \sin \eta \, {\bf e}_y}
                {|\bm{\tau}|} \,, \nn
\eea
where $|\bm{\tau}| \!=\! \sqrt{a^2-2ab \cos 2\eta + b^2}\,$. The ``rigid-body
rotation'' condition [Eq.\ (\ref{eq:rigidrotation})] can then be
written as 
\be
\label{eq:normal_ke}
U = \frac{ \Omega }{ |\bm{\tau}(\eta)| } \, 2 a b \, \sin 2 \eta \,.
\ee
In addition the distance between two points on the ellipse takes the
simple form
\bea
R^2 &\equiv& | {\bf r}(\eta) - {\bf r}(\eta') |^2 \\
&=&     4(a^2+b^2) \sin^2 \left(\frac{\eta-\eta'}{2} \right) 
        \left[1-\frac{2 a b}{a^2+b^2} \cos(\eta+\eta') \right] \!, \nn
\eea
and the dot product involved in Eqs.\ (\ref{eq:normalv})
and (\ref{eq:normalv-vp}) is given by
\be
{\bf n}(\eta) \cdot \bm{\tau}(\eta') =
        \frac{1}{|\bm{\tau}(\eta)|} (a^2-b^2) \sin(\eta -\eta') \,.
\ee

\end{multicols}
\subsection{Vortex patches---exact solution} \vspace{-.3cm}

It is now simple to show that an ellipse is, indeed, a uniformly
rotating shape for the vortex patch [Eqs.\ (\ref{eq:rigidrotation})
and (\ref{eq:normalv-vp})], since the normal velocity is given by
\bea
\label{eq:vp_ke1}
U_{\rm v} &=& -\frac{\omega_p}{4 \pi |\bm{\tau}(\eta)|} \, 
        \int_0^{2 \pi} \!\!\! d \eta' \, \ln \left\{
           \left[4(a^2+b^2) \sin^2(\frac{\eta-\eta'}{2})\right] 
\times  \left[1-\frac{2 a b}{a^2+b^2} \cos(\eta+\eta')\right]
        \right\}  \sin (\eta-\eta') \\
&=& -\frac{\omega_p}{4 \pi |\bm{\tau}(\eta)|}\,\int_0^{2 \pi} \!\!\! dx
        \, \ln \left[  1-\frac{2 a b}{a^2+b^2} \cos x \right] 
                \times \left(\sin 2 \eta \, \cos x - 
                \cos 2 \eta \, \sin x \right) \,,
\label{eq:vp_ke2}
\eea
%
\begin{multicols}{2}

\noindent
where in going from Eq.\ (\ref{eq:vp_ke1}) to (\ref{eq:vp_ke2}) we
eliminated the term inside the first braces in the logarithm due to
symmetry and changed variables to $x\!=\!\eta\!+\!\eta'$. Finally, the
term proportional to $\cos 2 \eta$ vanishes upon integration and we are
left with a simple integral, proportional to $\sin 2 \eta$. As long as
$a\!>\!b$ (for $a\!=\!b$ the ellipse has collapsed into a line), the
integral exists in closed form: 
\be
U_{\rm v} = {\omega_p} \, \frac{(a^2-b^2)b}{2 a |\bm{\tau}(\eta)|}
        \, \sin 2 \eta \,,
\ee
which, by direct comparison with Eq.\ (\ref{eq:normal_ke}), yields the
angular velocity
\be \Omega_{\rm v} = \frac{\omega_p}{4}\left(1
        -\frac{b^2}{a^2}\right)\,.
\ee

Consider now the parametrization given by Eq.\ (\ref{eq:paramet}) with
$b_l$ given by the first row in Table II. 
The maximum and minimum radii correspond to $R(\varphi)$ for $\varphi$
equal to 0 and $\pi/2$ respectively (for $b_2>0$). It is easy to see
that $a \!\sim\! 1 \!+\! 3 b_2^2 \!-\! b_2^4$ and 
$b \!\sim\! 2 b_2 \!+\! 5 b_2^3$, thus
$\Omega/\omega_p \!\simeq\! 1/2 \!-\! b_2^2 \!+\! b_2^4$, 
as shown in Table II. 
A simple Fourier analysis of Eq.\ (\ref{eq:radius_ke}) also results in
coefficients $b_l$ which agree with the perturbative solution.

\subsection{Two dimensional electron systems---no exact solution}
\vspace{-.3cm} 

It is also simple to see why an ellipse is {\em not} a stationary
solution of the two-dimensional electron system. Instead of the
logarithm, one has to deal with $1/\sqrt{R^2}$, and the elimination of
the first term inside the braces is not possible. The resulting
integrands depend on $\eta$ in a non-trivial way, and the normal
velocity is {\em not} proportional to 
$|{\bm{\tau}}(\eta)|^{-1} \, \sin 2 \eta$.

\section{Geometry of planar curve motion}  \vspace{-.3cm}
\label{sec:curves}

\begin{figure}
  \begin{center}
    \leavevmode
    \epsffile{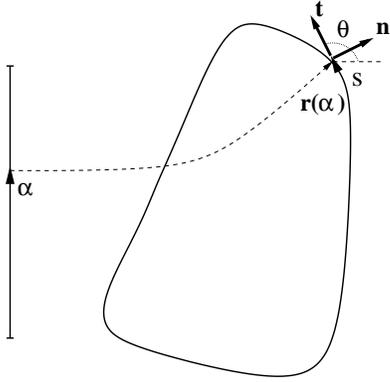}
  \end{center}
  \caption{ \label{fig:curve2d}
        Parametrization of a closed curve by a parameter $\alpha$
        defined on a fixed interval.  The illustration shows the
        arc-length $s$, the normal and tangential unit vectors 
        ${\bf n}$ and ${\bf t}$ and the tangent angle $\theta$.}
\end{figure}

For completeness, it is worth considering some general features of the
planar curve motion.\cite{Brower}  Consider a curve described by some
parametrization ${\bf r}(\alpha)$, where $\alpha$ is a parameter
defined on a fixed interval (see Fig.\ \ref{fig:curve2d}).  It is then
possible to consider tangent and normal unit vectors defined by
\be
\begin{array}{l}
\displaystyle
{\bf t} \equiv \frac{\bm{\tau}(\alpha)}{|\bm{\tau}(\alpha)|}\,,
        \;\;\;\; {\rm where } \;\;
        \bm{\tau}(\alpha) = {\bf r}_\alpha = 
        \frac{\partial {\bf r}(\alpha)} {\partial \alpha} \,, \\
\displaystyle
{\bf n} \equiv - {\bf e}_z \times {\bm t} \,.
\end{array} 
\ee
The arc-length $s(\alpha)$ corresponds to the length of the curve from
some arbitrary point to ${\bf r}(\alpha)$ and is defined by
$ds = \sqrt{dx^2 + dy^2} = \sqrt{g} \, d\alpha$, where the metric $g$
is defined by $g \equiv \bm{\tau \cdot \tau}$.  We then have the 
Frenet-Serret relations
\be
\begin{array}{l}
        {\bf r}_s = {\bf t} \,, \\
        {\bf t}_s = -\kappa \, {\bf n} \,,\\
        {\bf n}_s = \kappa \, {\bf t} \,,
        \end{array} 
\ee
which define the curvature $\kappa$.  It is also possible to define
the curve in terms of its tangent angle $\theta(s)$, where 
$\theta_s = \kappa$.

The kinematics of the curve can be determined if the velocity of the
curve ${\bf r}_t$ is determined.  It is convenient to decompose this
velocity into its {\rm normal} and {\em tangential} velocities (at
fixed $\alpha$):
\be
{\bf r}_t = U \, {\bf n} + W \, {\bf t}\,.
\ee
In general, $U$ and $W$ may be arbitrary functionals of ${\bf r}(s)$
and its derivatives.  It is easy to show that
\be 
\label{eq:misc_kin}
\begin{array}{l}
{\bf n}_t = - (U_s - \kappa W) {\bf t} \,,\\
{\bf t}_t = (U_s - \kappa W) {\bf n} \,,\\
(\sqrt{g})_t = W_\alpha + \sqrt{g}\, \kappa U \,.
\end{array} 
\ee

The length ${\cal L}$ and area ${\cal A}$ of the curve are given by
\bea
&& {\cal L} \equiv \oint ds = \int \!\! \sqrt{g} \, d\alpha \,, \\
&& {\cal A} = \frac{1}{2} \oint ({\bf r} \times {\bf t})
	\cdot {\bf e}_z \, ds= 
        \frac{1}{2} \int ({\bf r} \times {\bf r}_\alpha)
	\cdot {\bf e}_z  \, d\alpha \,,
\eea
and it is evident from the expressions above that their time
derivatives are equal to
\bea
\label{eq:Lt}
&& {\cal L}_t = \oint (\kappa U + W_s) \, ds = \oint \kappa U \, ds \,, \\
\label{eq:At}
&& {\cal A}_t = \oint U(s) \, ds \,,
\eea
where in the rightmost part of Eq.\ (\ref{eq:Lt}) we assumed that $W$
was periodic. 

Consider now the following identity 
\be
\frac{\partial}{\partial t}\frac{\partial}{\partial s} = 
\frac{\partial}{\partial t} \frac{1}{\sqrt{g}} 
\frac{\partial}{\partial \alpha} =
- \frac{g_t}{2 g^{3/2}} \frac{\partial}{\partial \alpha} +
\frac{\partial}{\partial s}\frac{\partial}{\partial t} \,.
\ee
Using Eq.\ (\ref{eq:misc_kin}), we obtain the commutator between the
arc-length and time derivatives
\be
\label{eq:commutator}
\left[ \frac{\partial}{\partial s},\frac{\partial}{\partial t} \right] = 
\left( W_s + \kappa U \right) \frac{\partial}{\partial s}\,.
\ee

We are finally able to determine a kinematic equation for the
curvature $\kappa$.  While the procedure is completely general, this
becomes particularly important when the dynamics obeys a geometric law
of motion, that is when $U$ and $W$ are functions of the curvature
and its derivatives only.  Using Eqs.\ (\ref{eq:misc_kin}) and 
(\ref{eq:commutator}) to calculate the time derivative of the
curvature at fixed $\alpha$ we find that
\be
\left[ \frac{\partial \kappa}{\partial t} \right]_\alpha = 
- [\kappa^2 + \partial_{ss} ] U + \kappa_s W \,.
\ee
It is convenient, however, to write a differential equation in a
parametrization-independent form, that is the time derivative should
be evaluated at fixed arc-length $s$.  Since $\partial_t|_s =
\partial_t|_\alpha - \kappa_s \int_0^s [\kappa U + W_{s'}] \, ds'$, we
have
\be
\label{eq:kappa_t}
\left[ \frac{\partial \kappa}{\partial t} \right]_s =
- [   \kappa^2 + \partial_{ss} ] \, U + \kappa_s W 
- \kappa_s \int_0^s [\kappa U + W_{s'}] \, ds' \,,
\ee
which is the form used in Eq.\ (\ref{eq:diffKappa}).  


\section{Elliptic integrals, elliptic functions and boundary
	conditions}  \vspace{-.3cm}
\label{sec:elliptic}

A quick inspection of Eq.\ (\ref{eq:integral}) reveals the following
possible alternative solutions, in terms of elliptic
functions,\cite{as} which depend on the nature of the zeros of $V(\kappa)$:

\noindent
\begin{enumerate}

\item
{Four real zeros
$\alpha \!\ge\! \kappa \!\ge\! \beta \!>\! \gamma \!>\! \delta$
(Fig. \ref{fig:pbc}-d):}

\bea
\hspace{-.8cm}
z (\kappa) &\!=\!&
	\frac{4}{\sqrt{(\alpha \!-\! \gamma)(\beta \!-\! \delta)}}  \,\!
        {\large F} \! \left( \! \sin^{-\!1} \!\! \!
        \sqrt{ \! \frac{(\alpha \!-\! \gamma)(\kappa \!-\! \beta)}
                {(\alpha\!-\!\beta)(\kappa\!-\!\gamma)} } ,\!
                        \lambda_4 \!\right) \!, \!\!\\
\hspace{-.8cm}
\lambda_4^2 &\!=\!& (\alpha - \beta)(\gamma - \delta)/ 
(\alpha - \gamma)(\beta - \delta) \,,
\eea
which can be inverted to read 
\be
\label{eq:solution4}
\kappa(z) = \frac{ \beta(\gamma\-\alpha) + \gamma(\alpha-\beta) 
        \, {\rm sn}^2 [ c_4 \, z, \lambda_4 ]}
        { (\gamma-\alpha) + (\alpha-\beta) 
        \, {\rm sn}^2 [ c_4 \, z, \lambda_4 ]} \,, 
\ee
with $c_4^2 \!=\! (\alpha \!-\! \gamma)(\beta \!-\!\delta)/16$. 
The period of $\kappa(z)$ is given in terms of the complete elliptic
integral $L_4 = (2/c_4) K (\lambda_4)$. 

\item
{Two real and two complex conjugate zeros $\alpha \!\ge\!
  \kappa \!\ge\! \beta$, $m\!\pm\!in$  (Fig. \ref{fig:pbc}-a,b,c):}

\bea
z (\kappa) &=& \frac{2}{\sqrt{p q}} \, {\large F} \! \left(
        2 \tan^{-1} \sqrt{ \frac{q(\alpha\!-\kappa)}{p(\kappa\!-\!\beta)}}, 
        \lambda_2 \right)\!, \\
p^2 &=& (m-\alpha)^2 + n^2 \,, \nn \\
q^2 &=& (m-\beta)^2 + n^2 \,, \\
\lambda_2^2 &=&  \left[(\kappa_{\rm max} - \kappa_{\rm min})^2 -
    (p-q)^2\right]/4 p q \,, \nn
\eea
which can be inverted to read
\be
\label{eq:solution2}
\kappa(z) = \frac{ (\beta p + \alpha q) + (\beta p - \alpha q)
        \, {\rm cn} [ \sqrt{pq} z/2 , \lambda_2 ]}
        { (p + q) + (p - q)
        \, {\rm cn} [  \sqrt{pq} z/2 , \lambda_2 ]} \,.
\ee
The period of $\kappa(z)$ is given in this case by by 
$L_2 = (8/\sqrt{pq}) K (\lambda_2)$.

\end{enumerate}

\begin{figure}
\begin{center}
\leavevmode
\epsfbox{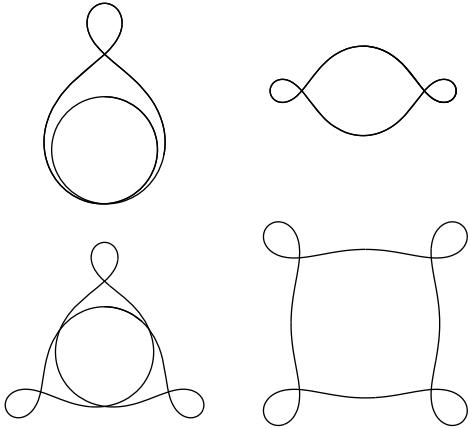}
\end{center}
\caption{\label{fig:4real}
  Examples of closed boundaries for the solutions corresponding to
  four real zeros of $V(\kappa)$.  Note that all curves cross
  themselves at least once and are not physical solutions for the
  problem of interest.}
\end{figure}

Since there is no cubic term in the potential $V(\kappa)$, the
sum of all roots must vanish. That leaves only one free parameter in
each case, once the minimum and maximum curvatures are fixed.  As
mentioned in Sec.\ \ref{sec:lia}, this free parameter is determined by
the boundary condition that the curve is closed and without
self-crossings.  
For the case of four real zeros, it is not possible to find solutions
that satisfy these conditions.  It is still possible to find beautiful
closed curves (Fig. \ref{fig:4real}), but these do not correspond to
physical solutions for the problem under consideration. 
The case with two real and two complex conjugate solutions does have
physical solutions and is discussed in detail in Sec.\ \ref{sec:lia}.  

\section{The straight infinite edge}  \vspace{-.3cm}
\label{sec:infinite}

In Sec.\ \ref{sec:lia} we discussed the invariant shapes of a closed
curve when the curvature satisfies a modified Korteweg-de Vries
equation (\ref{eq:mkdv}).  Single soliton solutions where the
curvature satisfied $\kappa(s,t) = \kappa(s-ct)$ were obtained by
twice integrating Eq.\ (\ref{eq:mkdv}), thus resulting in Eq.\
(\ref{eq:ode}), which can then be reduced to a simple problem in
quadratures [Eq.\ (\ref{eq:integral})] and the solutions for the
curvature in terms of Jacobi elliptic functions were discussed in that
context [see also Appendix \ref{sec:elliptic}]. 

In the case of an infinite curve, however, $\kappa$ and all its
arc-length derivatives should vanish as $s \rightarrow \pm \infty$,
and the constants of integration $a = b = 0$ in Eqs.\ (\ref{eq:ode})
and (\ref{eq:Vg}), so that $V(\kappa) = \kappa^4/4 - c \kappa^2$.  The
traveling wave solutions can then be found as
\be
\label{eq:inf-kappa}
\kappa (z) = 2 \sqrt{-c} \; {\rm sech} [ \sqrt{-c} \,(s - c t)] \,, 
\;\;\;\;\; c < 0 \,.
\ee
The tangent angle and the shape can then be obtained by a simple
integration:
\bea
\label{eq:inf-theta}
&&\theta (z) = \!\int \!\!\kappa \, ds = 4 \arctan \! \left\{\!
\tanh \!\left[\frac{\sqrt{-c}}{2}(s \!-\! c t) \!\right] \!\right\} +
\pi , \!\\
\label{eq:inf-Z}
&&x + i\,y = \int_0^s e^{i \theta} \, ds' \nn \\ 
&& \;\;= (s-ct) - \frac{4}{\sqrt{-c}} \left[ 
\frac{1}{i + \tanh[\!\frac{\sqrt{-c}}{2}(s - c t)]} - i\frac{\pi}{4} 
\right]\!.\!
\eea
Unfortunately these curves represent a small loop traveling along the
boundary, and self-crossing solutions are not possible 
for the boundary of a physical system.

\begin{figure}[h]
\begin{center}
\leavevmode
\epsfbox{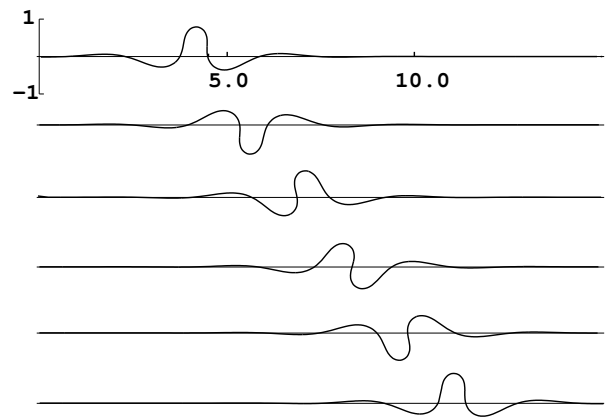}
\end{center}
\caption{\label{fig:breather}
  A breather soliton for $k = 2$ and $l=1$.  In this time sequence
  $\Delta t = 0.125$.}
\end{figure}

It is, however, possible to have more complicated solutions that have
a traveling envelope with time dependent oscillations within it.  One
such example is the ``breather'' solution,\cite{mKdV} which loosely
speaking, corresponds to a pair of bound solitons: 
\be
 \label{eq:breather-kappa}
 \kappa(s,t) = - 4 \frac{\partial}{\partial s} 
 \arctan \left\{ \frac{l}{k} \frac{ \sin[ k \, s - k (k^2 - 3 l^2) t]} 
   {\cosh [l \, s - l (3 k^2 - l^2) t]} \right\} ,
\ee
where $l$ and $k$ are arbitrary.  The value for the tangent angle in
this case is evident since $\theta_s = \kappa$; however, the shape of
the curve requires numerical integration.  Figure \ref{fig:breather}
shows a sequence corresponding to the motion of one example of a
breather.

%

\end{multicols}

\references

\bibitem{mechanics} See for example H. Goldstein, 
        {\em Classical Mechanics} (Addison-Wesley, Reading, 1980),
        2nd. Ed., Chap. 1,2 and 8.      
\bibitem{LLFM} L. D. Landau and E. M. Lifshitz, {\em Fluid Mechanics}
(Pergamon Press, Oxford, 1959) \S 12.
\bibitem{LiquidDropEXPT} R. G. Holt and E. H. Trinh,
        Phys. Rev. Lett. {\bf 77}, 1274 (1996);
  R. E. Apfel {\em et al.}, Phys. Rev. Lett. {\bf 78}, 1912 (1997).
\bibitem{LiquidDropMKDV} A. Ludu and J. P. Draayer,
        Phys. Rev. Lett. {\bf 80}, 2125 (1998).
\bibitem{non-neutral} T. M. O'Neil, Physics Today {\bf 52} vol. 2, 24
        (1999), and references therein.
\bibitem{Nuclear} A. Bohr, Mat. Fys. Medd. Dan. Vid. Selsk. {\bf 26}
  (1952) no. 14; A. Bohr and B. R. Mottelson, {\em Nuclear Structure},
  Vol. 1 (Benjamin, New York, 1969).
\bibitem{PlasmaClouds} E. A. Overman and N. J. Zabusky,
  Phys. Rev. Lett. {\bf 45}, 1693 (1980).
  E.A. Overman, N.J. Zabusky and S.L. Ossakow, Phys. Fluids {\bf 26},
  1139 (1983). 
\bibitem{PatternFormation} R. E. Rosensweig, 
        {\em Ferrohydrodynamics} (Cambridge Univ. Press, 1985); 
  S. A. Langer, R. E. Goldstein and D. P. Jackson, 
          Phys. Rev. A {\bf 46}, 4894 (1992); 
  A. T. Dorsey and R. E. Goldstein, Phys. Rev. B {\bf 57}, 3058 (1998).
\bibitem{lamb} Sir H. Lamb, {\em Hydrodynamics} (Dover, New York
        1932) Sect. 158 and 159.
\bibitem{ContourAlgorithm} N. J. Zabusky, M. H. Hughes and K. V. Roberts,
  J. Comp. Phys. {\bf 30}, 96 (1979); N. J. Zabusky and E. A. Overman,
  {\em ibid} {\bf 52}, 351 (1983).
\bibitem{VStates} G. S. Deem and N. J. Zabusky, 
        Phys. Rev. Lett. {\bf 40}, 859 (1978).
\bibitem{VortexPatchDynamics} R. E. Goldstein and D. M. Petrich,
        Phys. Rev. Lett. {\bf 69}, 555 (1992).
\bibitem{EdgesQHE}
  See X. G. Wen, Int. J. Mod. Phys. B {\bf 6}, 1711 (1992); and
  C. L. Kane and M. P. A. Fisher, in 
  {\em Perspectives in Quantum Hall Effects: Novel Quantum Liquids in
   Low-Dimensional Semiconductor Structures}, 
  edited by S. Das Sarma and A. Pinczuk (Wiley, New York, 1997),
  pp.\ 109-159. 
\bibitem{wexler99} C. Wexler and A. T. Dorsey, Phys. Rev. Lett. 
        {\bf 82}, 620 (1999).
\bibitem{mKdV} D. J. Korteweg and G. de Vries, 
  Phil. Mag. (5), {\bf 39}, 422 (1895); 
  A. Das, {\it Integrable Models} (World Scientific, Singapore, 1989);
  P. G. Drazin and R. S. Johnson, 
  {\em Solitons: an Introduction} (Cambridge University Press,
  Cambridge, 1989).
\bibitem{KdVHierarchy}  R. E. Goldstein and D. M. Petrich,
  Phys. Rev. Lett. {\bf 67}, 3203 (1991); in 
  {\em Singularities in Fluids, Plasmas and Optics} (Kluwer Acad.,
  1993), pp. 93-109. 
\bibitem{EMP} V. A. Volkov and S. A. Mikhailov, 
        Zh. Eksp. Teor. Fiz. {\bf 94}, 217 (1988) 
        [Sov. Phys. JETP {\bf 67}, 1639 (1988)]. 
\bibitem{linear} S. Giovanazzi, L. Pitaevskii, and S. Stringari,
	Phys. Rev. Lett. {\bf 72}, 3230 (1994);
	A. Cappelli, C. A. Trugenberger, and G. R. Zemba,
	Ann. Phys. {\bf 246}, 86 (1996).
\bibitem{winf}
	S. Iso, D. Karabali, and B. Sakita, Nucl. Phys. B {\bf 388}, 
	700 (1992);
	Phys. Lett. B {\bf 296}, 143 (1992);
	D. Karabali, Nucl. Phys. B {\bf 428}, 531 (1994).
\bibitem{glazman}  I. L. Aleiner and L. I. Glazman, 
        Phys. Rev. Lett. {\bf 72}, 2935 (1994).  
  I. L. Aleiner, D. Yue and L. I. Glazman, 
        Phys. Rev. B {\bf 51}, 13467 (1995).
\bibitem{AcousticModes} G. Ernst, 
  R.J. Haug, J. Khul, K. von Klitzing and K. Eberl, 
  Phys. Rev. Lett. {\bf 77}, 4245 (1996).
\bibitem{as} M. Abramowitz and I. Stegun, 
        {\em Handbook of Mathematical Functions}, (Dover, New York, 1965).
\bibitem{FreqDomain}  S. J. Allen, H. L. Stormer and J. C. M. Hwang, 
        Phys. Rev. B {\bf 28}, 4875 (1983); 
   M. Wassermeier {\em et al.}, Phys. Rev. B {\bf 41}, 10287 (1990); 
   I. Grodnesky, D. Heitman and K. von Klitzing, 
        Phys. Rev. Lett. {\bf 67}, 1019 (1991).      
\bibitem{TimeDomain} R. C. Ashoori H. L. Stormer, L. N. Pfeiffer, K. W.
	Baldwin, and K. West, Phys. Rev. B {\bf 45}, 3894 (1992). 
\bibitem{TimeDomain2} G. Ernst, N. B. Zhitenev, R. J. Haug, and K. von
	Klitzing, Phys. Rev. Lett. {\bf 79}, 3748 (1997). 
\bibitem{batchelor} See G. K. Batchelor, {\it An Introduction to Fluid
    Dynamics} (Cambridge University Press, Cambridge, 1967), Ch.\ 7. 
\bibitem{Brower} R. C. Brower, D. A. Kessler, J. H. Koplik and  
	J. H. Levine, Phys. Rev. A {\bf 29}, 1335 (1984).
\bibitem{conservation} J. Miller, Phys. Rev. Lett. {\bf 65}, 2137
	(1990); J. Miller, P. B. Weichman, and M. C. Cross,
	Phys. Rev. A {\bf 45}, 2328 (1992).
\bibitem{zhitenev} N. B. Zhitenev, R. J. Haug, K. v. Klitzing, and K.
	Eberl, Phys. Rev. B {\bf 52}, 11277 (1995).
\end{document}